# Accessing dark states optically through excitation-ferrying states

Zixuan Hu[1,2], Gregory S. Engel[3], Sabre Kais*[1,2]

1. Department of Chemistry, Department of Physics, and Birck Nanotechnology Center, Purdue University, West Lafayette, IN 47907, United States
2. Qatar Environment and Energy Research Institute, College of Science and Engineering, HBKU, Qatar
3. Department of Chemistry, University of Chicago, Chicago, IL 60637, United States

*Email: kais@purdue.edu

The efficiency of solar energy harvesting systems is largely determined by their ability to transfer excitations from the antenna to the energy trapping center before recombination. Dark state protection, achieved by coherent coupling between subunits in the antenna structure, can significantly reduce radiative recombination and enhance the efficiency of energy trapping. Because the dark states cannot be populated by optical transitions from the ground state, they are usually accessed through phononic relaxation from the bright states. In this study, we explore a novel way of connecting the dark states and the bright states via optical transitions. In a ring-like chromophore system inspired by natural photosynthetic antennae, the single-excitation bright state can be optically connected to the lowest energy single-excitation dark state through certain double-excitation states. We call such double-excitation states the ferry states and show that they are the result of accidental degeneracy between two categories of double-excitation states. We then mathematically prove that the ferry states are only available when $N$, the number of subunits on the ring, satisfies $N = 4l + 2$ ($l$ being an integer). Numerical calculations confirm that the ferry states enhance the energy transfer power of our model, showing a significant energy transfer power spike at $N = 6$ compared with smaller $N$ values, even without phononic relaxation. The proposed mathematical theory for the ferry states is not restricted to this one particular system or numerical model. In fact, it is potentially applicable to any coherent optical system that adopts a ring-shaped chromophore arrangement. Beyond the ideal case, the ferry state mechanism also demonstrates robustness under weak phononic dissipation, weak site energy disorder, and large coupling strength disorder.

*Introduction.* --- Reduction of the exciton recombination rate remains a fundamental limiting factor in designing high efficiency light harvesting systems[1]. Indeed, radiative recombination critically contributes to the Shockley–Queisser limit on photovoltaic energy conversion[2]. Pioneering work by Scully and coworkers[3,4] has shown that the detailed balance can be broken with coherence induced by either microwave radiation or by noise. Natural light harvesting complexes (LHC) commonly contain multiple chromophores that coherently couple to each other through dipole-dipole interactions[5-16]. In artificial models inspired by natural LHCs, the coherent coupling between chromophore subunits creates dark states which, when populated, prevent the excitons from radiatively recombining, effectively increasing the energy transfer efficiency of the LHC[17-19]. It is assumed that the dark states are accessed through phononic relaxation from the bright state, when the intraband eigenenergy spacing of the antenna Hamiltonian is comparable to the vibrational energy gaps of the environment. Indeed, when the phononic relaxation is turned off, the dark states are not populated and the increase in energy trapping efficiency is reduced[19,20].

Theoretical and experimental studies typically examine the single-excitation manifold, and the terms dark and bright states refer to the eigenstates of the Hamiltonian accessible from the ground state via a single excitation[21-23]. Indeed, under the solar illumination[17], the steady-state populations of the single-excitation eigenstates in photosynthetic antennae are generally orders of magnitude greater than those of the double-excitation states. Nevertheless, in our current study of the excitation transfer dynamics of multi-dipole systems under weak illumination, the double-excitation manifold significantly affects the transfer rate when the number of dipoles is greater than five. In particular we show that, despite the negligible steady-state population, certain double-excitation states greatly enhance exciton trapping by optically connecting the single-excitation bright states to the single-excitation dark states. We call such double-excitation states the ferry states and show they are the result of accidental degeneracy between two categories of double-excitation states. A characteristic of this perhaps surprising phenomenon is that the ferry states are unique to a ring-structured antenna system with $N = 4l+2$ chromophore subunits ($l$ being an integer). Thanks to the ferry states, a new channel, in addition to the phononic relaxation channel, is opened to transfer excitations to the dark states. This new channel greatly enhances the energy transfer power of our model light harvesting system. The highlight of this study is a mathematical proof for the availability of the ferry states, proposing a $N = 4l+2$ rule for the ring systems. The ability to optically access the dark states may have general implications for any optical system that adopts a ring-structured chromophore arrangement, which could lead to applications in a broader range of photonic devices.

*Models.* --- Our total system includes an antenna system that generates optical excitations and an energy trapping site that converts excitations received from the antenna into power. The antenna system consists of a ring of $N$ identical two-level optical emitters coupled through nearest-neighbor dipole-dipole interactions. The system's Hamiltonian is given by:

$$H_a = \omega \sum_{i=1}^{N} \sigma_i^+ \sigma_i^- + S \sum_{i=1}^{N} (\sigma_i^+ \sigma_{i+1}^- + \sigma_{i+1}^+ \sigma_i^-) \quad (1)$$

where $\hbar = 1$, $\omega$ is the site energy, $S$ is the coupling strength, and $\sigma_{N+1}^\pm = \sigma_1^\pm$. The use of the Pauli raising and lowering operators explicitly ensures that a single site cannot support more than one excitation. Under Jordan-Wigner transformation, the Hamiltonian becomes:

$$H_{JW} = \omega \hat{n} + S \sum_{j=1}^{N-1} \left( c_j^\dagger c_{j+1} + c_{j+1}^\dagger c_j \right) - S \left( c_N^\dagger c_1 + c_1^\dagger c_N \right) e^{i\pi \hat{n}} \quad (2)$$

where $\hat{n}$ is the number operator for the total number of excitations on the ring and $c_j^\dagger$ and $c_j$ represent the hardcore bosonic creation and annihilation operators, respectively, that observe the canonical anticommutation relations for fermions. The solution to equation (2) is dependent on the parity of the excitation number, due to the $e^{i\pi \hat{n}}$ term in the boundary condition. For a given excitation number, n, we first find the creation operators for the single excitation component states (referred to as the component states in the following):

$$C_m^+ = \frac{1}{\sqrt{N}} \sum_{j=1}^{N} e^{i\frac{2m\pi}{N}j} c_j^\dagger \quad \text{if } n \text{ is odd}$$

$$C_m^+ = \frac{1}{\sqrt{N}} \sum_{j=1}^{N} e^{i\frac{(2m+1)\pi}{N}j} c_j^\dagger \quad \text{if } n \text{ is even} \tag{3}$$

where $m$ ranges from zero to $N-1$. We then proceed to construct the n-excitation eigenstates by selecting n number of $C_m^+$ operators to operate successively on the ground state:

$$\left|\psi_{m_1 m_2 \ldots m_n}\right\rangle = C_{m_1}^+ C_{m_2}^+ \ldots C_{m_n}^+ \left|0\right\rangle \tag{4}$$

where $m_1 \neq m_2 \neq \ldots \neq m_n$. The energies of the n-excitation eigenstates are given by

$$\varepsilon_{m_1 m_2 \ldots m_n} = \sum_{i=1}^{n}\left(\omega + 2S\cos\frac{2m_i \pi}{N}\right) \text{ if n is odd and } \varepsilon_{m_1 m_2 \ldots m_n} = \sum_{i=1}^{n}\left(\omega + 2S\cos\frac{(2m_i + 1)\pi}{N}\right) \text{ if n is even,}$$

where the total energy is the sum of the energies of the individual component states. The single-excitation eigenstates are simply given by:

$$\left|\psi_m\right\rangle = \frac{1}{\sqrt{N}} \sum_{j=1}^{N} e^{i\frac{2m\pi}{N}j} c_j^\dagger \left|0\right\rangle \tag{5}$$

The antenna system is connected to the trapping site through the $N^{th}$ dipole. To calculate the steady state dynamics of the antenna-trap combined system, we follow the procedures used in Ref.[19,20] by setting up a standard Lindblad optical master equation:

$$\dot{\rho} = -i\left[H_a + H_t, \rho\right] + D_o\left[\rho\right] + D_p\left[\rho\right] + D_t\left[\rho\right] + D_x\left[\rho\right] \tag{6}$$

where $\rho = \rho_a \otimes \rho_t$ is the total density operator of the antenna-trap system, $H_t = \omega_t \sigma_t^+ \sigma_t^-$ is the trapping site Hamiltonian, $D_o\left[\rho\right]$ is the optical dissipator describing the interband transitions between different excitation levels, $D_p\left[\rho\right]$ is the phononic dissipator describing the intraband transitions within one excitation level, $D_t\left[\rho\right]$ describes the decay process of the trapping site, and $D_x\left[\rho\right]$ describes the extraction process from the antenna ring to the trapping site. As outlined later in this Letter, we initially remove the contribution from $D_p\left[\rho\right]$ to develop the theory with optical transitions only and build solid understanding of the ideal case, before proceeding to discuss scenarios involving weak phononic transition and disorder. Steady state solution of $\rho$ in equation (6) is obtained by the open-source quantum dynamics software QuTiP[24]. The quantum heat engine theory[4,25] is then used to calculate the power output of our light harvesting system. Details of the numerical model can be found in the Supplementary Information or Refs.[19,20].

*Emergence of the ferry states.* --- By simple algebraic evaluation of equation (5), only the $\left|\psi_0\right\rangle = \frac{1}{\sqrt{N}} \sum_{j=1}^{N} c_j^\dagger \left|0\right\rangle$ state has finite coupling to the ground through optical transitions:

$$\Gamma_{mg} = \left| \langle \psi_m | \sum_{i=1}^{N} \sigma_i^+ | 0 \rangle \right|^2 = N \delta_{m0},$$ where $\delta_{m0}$ is the Kronecker delta. Phononic dissipation is turned off, disallowing intraband transitions; therefore, we expect that only the totally symmetric $|\psi_0\rangle$ will be populated and that the energy transfer power will monotonically decrease over increasing dipole number $N$ due to the delocalization of the $|\psi_0\rangle$ excitation away from the trapping site. The numerical results are shown in Figure 1.

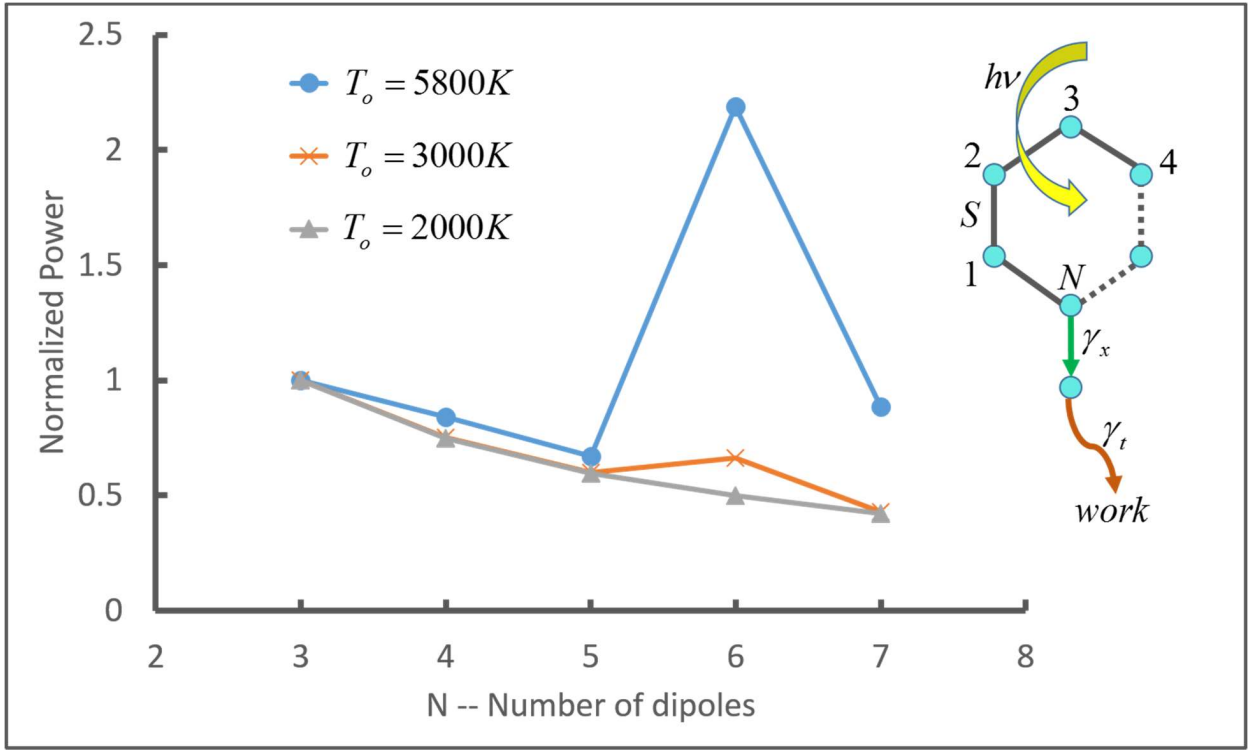

*Figure 1. Normalized power calculated with no intraband transition under different optical temperatures: 5800K, 3000K and 2000K. The insert on the right shows the model structure. Site energy $\omega = 1.76 eV$, coupling strength $S = 0.02 eV$, optical transition rate $\gamma_0 = 10^{-6} eV$, phononic transition rate $\gamma_p = 0$. For other parameters see the Supplementary Information.*

Figure 1 shows the normalized power versus the number of dipoles for the optical temperatures of 5800K, 3000K, and 2000K. For the solar temperature $T_o = 5800K$, the delocalization effect induces a decrease in power from $N = 3$ to $N = 5$, as expected. However, at $N = 6$, there is a significant power spike, which is unexpected if we consider only the single-excitation states. Here, the double-excitation eigenstates, although less populated by orders of magnitudes, play a surprising role in the energy transfer dynamics. For a double-excitation state to be populated, it must have finite optical coupling to the bright state $|\psi_0\rangle = \frac{1}{\sqrt{N}} \sum_{j=1}^{N} c_j^\dagger |0\rangle$. The energy ladder structure of the ring systems for $N = 3$ to $N = 6$ enables us to study these double-excitation states (Figure 2).

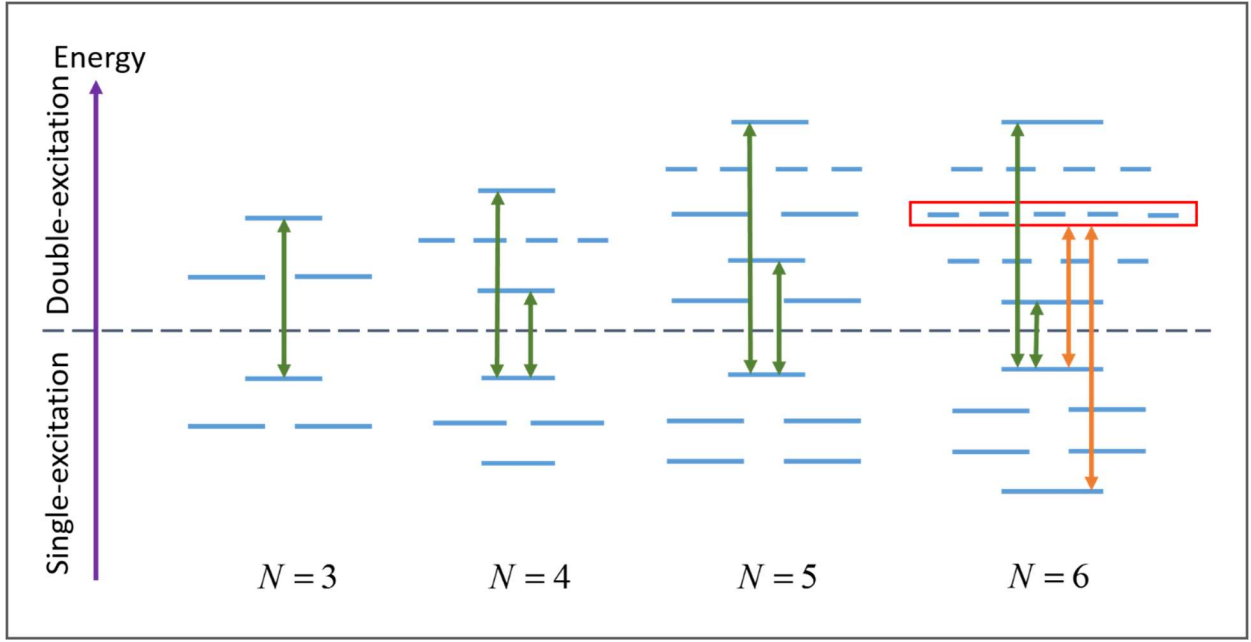

*Figure 2. Energy ladder structure of the single-excitation and double-excitation eigenstates for the ring of dipoles with N = 3 to N = 6 subunits. Single-excitation and double-excitation states are shown below and above the dashed line, respectively. The red box highlights the ferry states. Double-headed arrows mark the possible optical transition paths from the single-excitation bright states. The paths associated with the ferry states are shown in orange.*

In Figure 2, the highest energy single-excitation state is always the bright one, and its finite couplings to the double-excitation states are marked by double-headed arrows. For $N = 3$ to $N = 5$, the only double-excitation states with finite coupling to the bright single-excitation state are the nondegenerate states that do not couple to any other single-excitation states. However, on the $N = 6$ energy ladder, there is a level with five degenerate double-excitation states that couple to both the bright single-excitation state and the lowest dark single-excitation state (transition paths marked by orange double-headed arrows in Figure 2). This is an important property of the $N = 6$ energy ladder since these five double-excitation states allow the dark state to be populated from the bright state through optical transitions. We name these states the ferry states as they ferry excitations from the bright state to the dark state. Consequently, for $N = 6$, the energy transfer power is greatly enhanced by dark state protection, producing the spike observed in Figure 1. The optical temperature $T_o$ controls the population statistics of interband transitions, therefore lowering $T_o$ should reduce the population on the double-excitation states and reduce the effect of the ferry states. With this in mind, we expect the power spike to reduce and eventually disappear as $T_o$ is lowered. In Figure 1 we indeed observe the N = 6 spike disappear as the optical temperature decreases, supporting our claim that ferry states enhance the energy transfer power in our model light harvesting system.

*Theory of the ferry states.* --- To further investigate the ferry state phenomenon, we consider how the double-excitation states are formed from the component states with $n = 2$, as defined in equation (3). The energy ladder structure of the component states is shown in Figure 3:

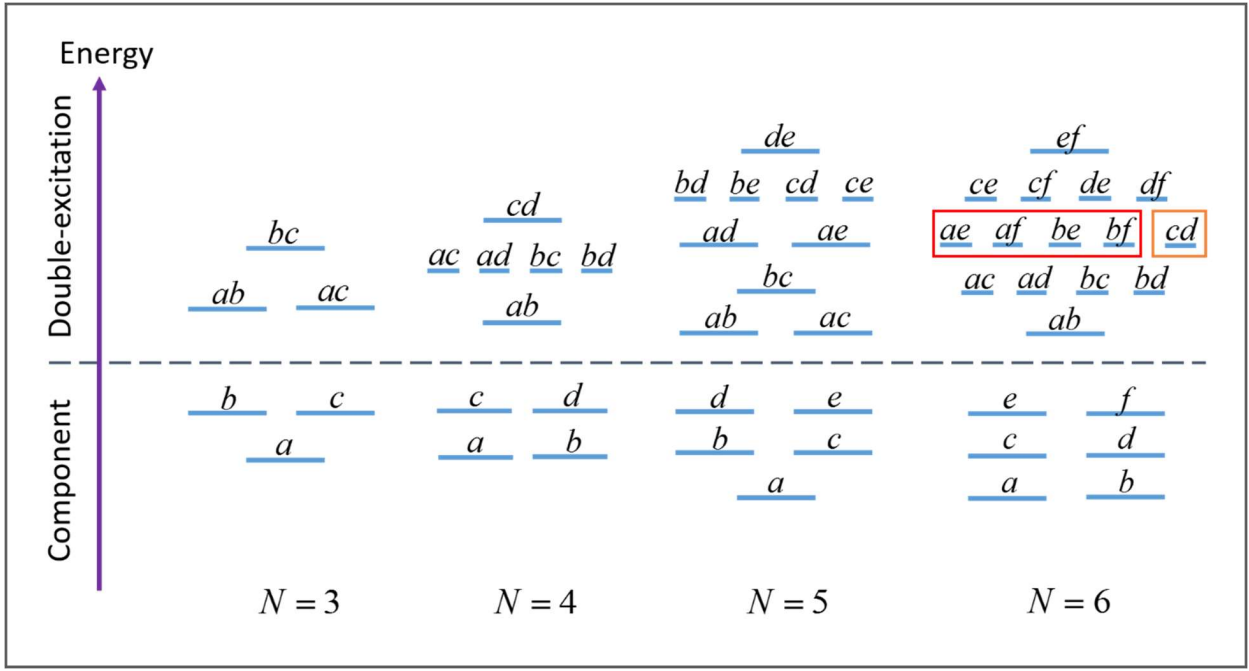

*Figure 3. Energy ladder structure showing how the double-excitation states are formed by the component states. Below the dashed line, the component states are labeled by single letters. Above the dashed line, the double-excitation states are labeled by double letters corresponding to their respective component states. The ferry states are separated into two categories – one formed by component states of different energies (first category; red box) and one formed by component states of the same energy (second category; orange box).*

In Figure 3, the nondegenerate double-excitation states (e.g. $|bc\rangle$ in $N = 3$, $|ab\rangle$ in $N = 6$, etc.) – states that couple to the bright single-excitation states in Figure 2 – are always formed by the doubly degenerate component states. On the other hand, the double-excitation states with even degeneracies (e.g. $|ab\rangle$ and $|ac\rangle$ in $N = 3$, $|ac\rangle$ level in $N = 4$, etc.) are always formed by the component states of different energies. The ferry state level of degeneracy five for $N = 6$ results from the accidental degeneracy between a collection of double-excitation states with degeneracy four and the state $|cd\rangle$. The accidental degeneracy is created when three energy levels are evenly spaced in the component energy ladder such that the sum of the energies of $|c\rangle$ and $|d\rangle$ is equal to the sum of the energies of $|a\rangle$ and $|e\rangle$. When does the component energy ladder have such a feature? The energy of a component state is given by $\varepsilon_m = \omega + 2S\cos\frac{(2m+1)\pi}{N}$, and the energy ladder of the component states is symmetric with respect to $\frac{(2m+1)\pi}{N} = \pi$. Without loss of generality, we focus only on the states with $\frac{(2m+1)\pi}{N} \leq \pi$, and find this part of the energy ladder is antisymmetric with respect to $\frac{(2m+1)\pi}{N} = \frac{\pi}{2}$. Hence, the three evenly spaced levels on the component energy ladder can exist if and only if there is a particular component state with $\frac{(2m+1)\pi}{N} = \frac{\pi}{2}$ (see the Supplementary Information for a formal proof). The condition for the accidental degeneracy is:

$$N = 4l+2 \qquad l=1,2,3,\ldots \tag{7}$$

Using this general formula supporting the accidental degeneracy shown for $N = 6$ in Figure 3, we proceed to prove that: 1) double-excitation states without the accidental degeneracy are never ferry states and 2) ferry states can always be formed from the double-excitation states with accidental degeneracy. We start by analytically formulating the coupling between the double-excitation states and the single-excitation states. From equations (3) and (4), a general double-excitation state is:

$$\left|\psi_{m_1 m_2}\right\rangle = C_{m_1}^+ C_{m_2}^+ \left|0\right\rangle = \frac{1}{N}\sum_{j<h}\left(e^{i\frac{\pi}{N}[(2m_1+1)j+(2m_2+1)h]} - e^{i\frac{\pi}{N}[(2m_1+1)h+(2m_2+1)j]}\right)\sigma_j^+\sigma_h^+\left|0\right\rangle \tag{8}$$

where we have carried out the Jordan-Wigner transformation. The coupling of $\left|\psi_{m_1 m_2}\right\rangle$ to a general single-excitation state is:

$$\begin{aligned}\Gamma &= \left|\left\langle\psi_{m_1 m_2}\middle|\sum_{j=1}^{N}\sigma_j^+\middle|\psi_k\right\rangle\right|^2 \\ &= \frac{1}{N^3}\left|\sum_{j<h}\left(e^{i\frac{\pi}{N}[(2m_1+1)j+(2m_2+1)h]} - e^{i\frac{\pi}{N}[(2m_1+1)h+(2m_2+1)j]}\right)\left(e^{i\frac{2k\pi}{N}h} + e^{i\frac{2k\pi}{N}j}\right)\right|^2\end{aligned} \tag{9}$$

If $\left|\psi_{m_1 m_2}\right\rangle$ is formed by $\left|C_{m_1}\right\rangle$ and $\left|C_{m_2}\right\rangle$ of different energies, it's coupling to the bright state is calculated by setting $\frac{\pi}{N}\left[(2m_1+1)+(2m_2+1)\right] \neq 2\pi$ and $k = 0$ in equation (9):

$$\Gamma_1 = \frac{4}{N^3}\left|\sum_{j<h}\left(e^{i\frac{\pi}{N}[(2m_1+1)j+(2m_2+1)h]} - e^{i\frac{\pi}{N}[(2m_1+1)h+(2m_2+1)j]}\right)\right|^2 = 0 \tag{10}$$

which is evaluated to be zero. The detailed mathematical derivations for this equation, and all following equations, can be found in the Supplementary Information. On the other hand, if $\left|\psi_{m_1 m_2}\right\rangle$ is formed by $\left|C_{m_1}\right\rangle$ and $\left|C_{m_2}\right\rangle$ of the same energy, $\frac{\pi}{N}\left[(2m_1+1)+(2m_2+1)\right] = 2\pi$, and the coupling to the bright state is given by:

$$\Gamma_2 = \frac{4}{N}\left(\frac{\sin\frac{(2m_2+1)\pi}{N}}{1-\cos\frac{(2m_2+1)\pi}{N}}\right)^2 \neq 0 \tag{11}$$

where the inequality to zero is obtained by noting $\frac{\pi}{N}[(2m_1+1)+(2m_2+1)] = 2\pi$ and $m_1 \neq m_2$ (recall equation (4)), such that $\frac{(2m_2+1)\pi}{N} \neq \pi$. Equation (10) means that if $|\psi_{m_1 m_2}\rangle$ is formed by $|C_{m_1}\rangle$ and $|C_{m_2}\rangle$ of different energies, it cannot couple to the bright state. On the other hand, equation (11) means if $|\psi_{m_1 m_2}\rangle$ is formed by $|C_{m_1}\rangle$ and $|C_{m_2}\rangle$ of the same energy, it couples to the bright state. Now, how about the coupling between the double-excitation states and the dark states? For $|\psi_{m_1 m_2}\rangle$ formed by $|C_{m_1}\rangle$ and $|C_{m_2}\rangle$ of the same energy, the coupling is calculated by setting $\frac{\pi}{N}[(2m_1+1)+(2m_2+1)] = 2\pi$ and $k \neq 0$ in equation (9):

$$\Gamma_3 = \frac{1}{N^3} \sum_{j<h} \left| \begin{array}{l} e^{i\frac{\pi}{N}[-(2m_2+1)j+(2m_2+2k+1)h]} - e^{i\frac{\pi}{N}[-(2m_2+1)h+(2m_2+2k+1)j]} \\ + e^{i\frac{\pi}{N}[-(2m_2-2k+1)j+(2m_2+1)h]} - e^{i\frac{\pi}{N}[-(2m_2-2k+1)h+(2m_2+1)j]} \end{array} \right|^2 = 0 \quad (12)$$

which means $|\psi_{m_1 m_2}\rangle$ will not couple to any dark state.

At this point, we are ready to mathematically prove that the double-excitation states without the accidental degeneracy are never ferry states. The double-excitation states can be partitioned into two categories: those formed by component states of different energies and those formed by component states of the same energy. By equation (10), the former category does not couple to the bright state and by equation (12), the latter does not couple to the dark states. Since there is no accidental degeneracy between the first category and the second category, they cannot form hybrid eigenstates by linear combination; therefore, no double-excitation eigenstates can couple to both the bright and dark states, Q.E.D.

To prove that ferry states can always be formed from the double-excitation states with accidental degeneracy, consider the following double-excitation state: $|\phi\rangle = \frac{1}{N} \sum_{j<h} \left( e^{i\pi j} e^{i\frac{\pi}{N}(h-j)} - e^{i\pi h} e^{i\frac{\pi}{N}(j-h)} \right) \sigma_j^+ \sigma_h^+ |0\rangle$ formed by $|C_{m_1}\rangle$ and $|C_{m_2}\rangle$ with $m_1 = \frac{N-2}{2}$ and $m_2 = 0$, and a particular dark state $|\psi_{m=N/2}\rangle = \frac{1}{\sqrt{N}} \sum_{j=1}^{N} (-1)^j c_j^\dagger |0\rangle$. The coupling between these two states is:

$$\Gamma_4 = \left| \langle \phi | \sum_{j=1}^{N} \sigma_j^+ | \psi_{m=N/2} \rangle \right|^2 = \frac{4}{N^3} \left| \sum_{j<h} \sin\frac{(h-j)\pi}{N} \left(1+(-1)^{j+h}\right) \right|^2 \neq 0 \quad (13)$$

where the inequality holds since $0 < \frac{(h-j)\pi}{N} < \pi$ such that $\sin\frac{(h-j)\pi}{N} > 0$. The energy of $|\phi\rangle$ is $\varepsilon_{m_1 m_2} = \sum_{i=1}^{2}\left(\omega + 2S\cos\frac{(2m_i+1)\pi}{N}\right) = 2\omega$ which coincides with the energy of the double-

excitation state formed by two equal energy component states: $|\chi\rangle = C_{m_3}^+ C_{m_4}^+ |0\rangle$ for which $\frac{(2m_3+1)\pi}{N} = \frac{\pi}{2}$ and $\frac{(2m_4+1)\pi}{N} = \frac{3\pi}{2}$. Therefore, $|\phi\rangle$ and $|\chi\rangle$ have accidental degeneracies between two categories of double-excitation states. Note that for these specific $m_3$ and $m_4$ to exist, $N$ must satisfy equation (7). By equation (13), the particular choice of $|\phi\rangle$ couples to the dark state $|\psi_{m=N/2}\rangle$ and by equation (11) the particular choice of $|\chi\rangle$ couples to the bright state. Since $|\phi\rangle$ and $|\chi\rangle$ are degenerate, a linear combination can be formed $|\psi_{hyb}\rangle = \alpha|\phi\rangle + \beta|\chi\rangle$. This linear combination is a double-excitation eigenstate of the system that couples to both the bright state and a dark state. Therefore, we have successfully constructed a ferry state when there is accidental degeneracy, Q.E.D.

*Robustness of the ferry state effect.* --- So far we have established a formal theory for the ferry state phenomenon under ideal conditions without phononic dissipation or disorder. When there is phononic dissipation, intraband transitions between single-excitation states will compete with interband transitions in mediating excitation transfer from the bright state to the dark states. In addition, intraband transitions between double-excitation states will cause the ferry states to lose excitations to other levels, weakening the ferrying mechanism. As a result, we expect that the ferry state effect is only visible when the phononic dissipation rate $\gamma_p$ is low compared with the optical transition rate $\gamma_o$. In such a case, the sign of the coupling strength, $S$, becomes important. If $S < 0$, the bright state has the lowest energy and is favored by thermal statistics, thereby making the ferry state effect more important in transferring excitation from the bright state to the dark states. Indeed, when $S < 0$ at a low phononic dissipation rate of $\gamma_p = 0.01\gamma_o$, we are able to observe the ferry state effects (see Supplementary Information). When $S > 0$, the dark states are favored by thermal statistics and the ferry state effect is not observed, even at $\gamma_p = 0.01\gamma_o$. Next, we consider two types of disorder possible in the system Hamiltonian in equation (1): one in the coupling strength $S$ and one in the site energy $\omega$. The ferry state degeneracy between two categories of double-excitation states is critical to the ferry state mechanism: to understand how the ferry state effect is affected by disorder, we need to study how the degeneracy is affected by disorder. Under disorder in the coupling strength $S$, the ferry state degeneracy is exactly preserved for $N = 4l + 2$ with any arbitrary choice of the individual $S$ values. The remarkable preservation of the degeneracy can be proven by analyzing the characteristic equation of the Hamiltonian with random $S$ values (see Supplementary Information). With this preservation of degeneracy, the ferry state effect is quite robust with coupling strength disorder and can be observed when the disorder is relatively large -- Gaussian standard deviation at 5% of the mean. For disorder in the site energy $\omega$, the ferry state degeneracy is preserved up to the first order perturbation energy. Consequently, the ferry state effect is less robust under site energy disorder and can be observed when the disorder is relatively small -- Gaussian standard deviation at 0.1% of the mean (see Supplementary Information). Overall, we see that the ferry state effect is robust under weak phononic dissipation, small $\omega$ disorder, and large $S$ disorder.

*Conclusions.* --- In this letter, we have studied the energy transfer dynamics of a system consisting of a ring-structured antenna connected to a trapping site. We have discovered that, in the absence of intraband transitions, there is an unexpected spike in the energy transfer power when there are $N = 6$ antenna emitters. The cause of the spike is the unique existence of the ferry states – double-excitation states that couple to both the single-excitation bright state and a single-excitation dark state – for $N = 6$. We confirmed that the ferry states enhanced the energy transfer power by lowering the optical temperature and observing the disappearance of the spike. We then characterized the ferry states and demonstrated that they are due to accidental degeneracy between two categories of double-excitation states. Furthermore, this accidental degeneracy was only available when $N = 4l + 2$. The ferry state effect is robust under weak phononic dissipation, small $\omega$ disorder, and large $S$ disorder. Mathematical proofs have been provided for the statements:
1. Double-excitation states without the accidental degeneracy are never ferry states, and
2. Ferry states can always be formed from double-excitation states with accidental degeneracy.
In conclusion, we have discovered and explained a novel mechanism to optically access the dark state from the bright state. This mechanism utilizes the ferry states available to the $N = 4l + 2$ membered ring structures. The ferry state mechanism and our analytic theory are not unique to the current model system; they are generally applicable to any coherent optical system described by a ring-structure Hamiltonian, as in equation (1). Promising applications may arise by designing materials that satisfy the $N = 4l + 2$ rule to maximize energy transfer power through a light harvesting system.

*Acknowledgement.* --- This work is supported by the QNRF exceptional grant: NPRPX-107-1-027

# Supplementary information: Accessing dark states optically through excitation-ferrying states

Zixuan Hu[1,2], Gregory S. Engel[3], Sabre Kais*[1,2]

4. Department of Chemistry, Department of Physics, and Birck Nanotechnology Center, Purdue University, West Lafayette, IN 47907, United States
5. Qatar Environment and Energy Research Institute, College of Science and Engineering, HBKU, Qatar
6. Department of Chemistry, University of Chicago, Chicago, IL 60637, United States

*Email: kais@purdue.edu

This supplementary document supports the discussion in the main text by providing theoretical, numerical, and technical details. Section 1 presents the numerical model. Section 2 gives the formal proof for the ferry state degeneracy condition $N = 4l + 2$. Section 3 details the mathematical derivations for the sufficient and necessary conditions of the ferry state effects. Section 4 provides the numerical results for non-ideal cases involving dissipation and disorder.

1. **Numerical model**

The antenna system is connected to the trapping site through the $N^{th}$ dipole. To calculate the steady state dynamics of the antenna-trap system, we follow the procedures used in [1,2] by writing the Lindblad optical master equation:

$$\dot{\rho} = -i[H_a + H_t, \rho] + D_o[\rho] + D_p[\rho] + D_t[\rho] + D_x[\rho] \qquad (1.1)$$

where $\rho = \rho_a \otimes \rho_t$ is the total density operator of the antenna-trap system and $H_t = \omega_t \sigma_t^+ \sigma_t^-$ is the trapping site Hamiltonian. $D_o[\rho] = \gamma_o \sum_{\omega_o} \Gamma_{K,K'} \left( N(\omega_o + 1) \mathcal{D}[\hat{L}_o, \rho] + N(\omega_o) \mathcal{D}[\hat{L}_o^\dagger, \rho] \right)$ is the optical dissipator describing the interband transitions between different excitation levels, where $\gamma_o$ gives the optical transition rate for the antenna, $\Gamma_{K,K'} = \left| \langle \psi_K | \sum_{j=1}^{N} \sigma_j^+ | \psi_{K'} \rangle \right|^2$ is the optical coupling strength between two eigenstates of the antenna Hamiltonian with $\omega_o = \varepsilon_K - \varepsilon_{K'} > 0$, $N(\omega_o) = \left( e^{\omega_o / k_B T_o} - 1 \right)^{-1}$ is the optical distribution, and $\mathcal{D}[\hat{L}_o, \rho] = \hat{L}_o \rho \hat{L}_o^\dagger - \frac{1}{2} \{ \hat{L}_o^\dagger \hat{L}_o, \rho \}$ is the Lindblad dissipator with $\hat{L}_o^\dagger = |K\rangle\langle K'|$. $D_p[\rho] = \gamma_p \sum_{\omega_p} \left( N(\omega_p + 1) \mathcal{D}[\hat{L}_p, \rho] + N(\omega_p) \mathcal{D}[\hat{L}_p^\dagger, \rho] \right)$ is the phononic dissipator describing the intraband transitions within one excitation level, where $\gamma_p$ gives the phononic relaxation rate, $N(\omega_p) = \left( e^{\omega_p / k_B T_p} - 1 \right)^{-1}$ is the thermal distribution, and $\mathcal{D}[\hat{L}_p, \rho] = \hat{L}_p \rho \hat{L}_p^\dagger - \frac{1}{2} \{ \hat{L}_p^\dagger \hat{L}_p, \rho \}$ and $\hat{L}_p^\dagger = |\mu\rangle\langle\nu|$ are the intraband transitions with $\omega_p = \varepsilon_\mu - \varepsilon_\nu > 0$. In this study, we initially remove the contribution from $D_p[\rho]$ by setting $\gamma_p = 0$ to focus on optical transitions only. $D_t[\rho] = \gamma_t \mathcal{D}[\sigma_t^-, \rho]$ describes the decay process of the trapping site, and $D_x[\rho] = \gamma_x \mathcal{D}[\sigma_N^- \sigma_t^+, \rho]$ describes the extraction process from the antenna ring to the trapping site. The parameters associated with each of the processes in equation (6) are given in Table 1:

| Parameter | Symbol | Value |
| --- | --- | --- |
| Antenna site energy | $\omega$ | 1.76 eV |
| Antenna coupling strength | $S$ | 0.02 eV |
| Antenna optical decay rate | $\gamma_o$ | $10^{-6}$ eV |
| Antenna phononic decay rate | $\gamma_p$ | set to zero |
| Trap optical decay rate | $\gamma_t$ | $0.1 \gamma_0$ |
| Antenna to trap extraction rate | $\gamma_x$ | $0.01 \gamma_0$ |
| Ambient temperature | $T_p$ | 300K |
| Optical temperature | $T_o$ | 5800K |

*Table 1. Parameters used in the numerical calculations.*

Steady state solution of $\rho = \rho_a \otimes \rho_t$ in equation (6) is obtained by the open-source quantum dynamics software QuTiP [3]. The theory of quantum heat engines [4,5] is then used to calculate the power output of our light harvesting system, for which the current is given by $I = e\gamma_t \langle \rho_{te} \rangle_{ss}$, the voltage is given by $eV = \hbar\omega_t + k_B T_p \ln\left(\frac{\langle \rho_{te} \rangle_{ss}}{\langle \rho_{tg} \rangle_{ss}}\right)$, where $\langle \rho_{te} \rangle_{ss}$ is the steady state population of the trap's excited state, $\langle \rho_{tg} \rangle_{ss}$ is the steady state population of the trap's ground state, $e$ is the fundamental charge, $k_B$ is the Boltzmann constant, and $T_p$ is the thermal temperature.

## 2. Formal proof for the ferry state degeneracy condition

In the main text, we have provided a descriptive argument that if, and only if, there is a particular component state with $\frac{(2m+1)\pi}{N} = \frac{\pi}{2}$ will we have the three evenly spaced levels on the component energy ladder, thereby creating the condition for the ferry state degeneracy: $N = 4l + 2$. In this section, we provide a formal proof for this statement. First, we abstract the statement into the following proposition:

*Proposition: Consider the function* $f(m) = \cos\left(\dfrac{2m+1}{N}\pi\right)$ *where the integer $m$ satisfies* $0 \leq m < N$ *and the integer* $N > 2$. *The situation* $f(m_1) - f(m_2) = f(m_2) - f(m_3)$ *while* $f(m_1) > f(m_2) > f(m_3)$ *can happen if, and only if,* $\dfrac{2m_2 + 1}{N} = \dfrac{1}{2}$ *or* $\dfrac{2m_2 + 1}{N} = \dfrac{3}{2}$ *and* $m_1 - m_2 = m_2 - m_3$.

Clearly, $f(m) = \cos\left(\dfrac{2m+1}{N}\pi\right)$ is the energy of the component states, and for the double-excitation states to have accidental degeneracy between the two categories, we must have three energy levels of the component states equally spaced as in $f(m_1) - f(m_2) = f(m_2) - f(m_3)$. The proposition outlined above provides a necessary and sufficient condition for this to happen. The $N = 4l + 2$ rule for the ferry states then follows.

Proof: In Conway and Jones, Acta Arith. XXX (1976) 229-240, Theorem 7 (abbreviated to CJ7 in the following) states that:

> Suppose we have at most four distinct rational multiples of $\pi$ lying strictly between $0$ and $\pi/2$ for which some rational linear combination of their cosines is rational but no proper subset has this property. Then the appropriate linear combination is proportional to one from the following list:

$\cos \pi/3 = 1/2$,
$-\cos \phi + \cos(\pi/3 - \phi) + \cos(\pi/3 + \phi) = 0$,
$\cos \pi/5 - \cos 2\pi/5 = 1/2$,
$\cos \pi/7 - \cos 2\pi/7 + \cos 3\pi/7 = 1/2$,
$\cos \pi/5 - \cos \pi/15 + \cos 4\pi/15 = 1/2$,
$-\cos 2\pi/5 + \cos 2\pi/15 - \cos 7\pi/15 = 1/2$,
$\cos \pi/7 + \cos 3\pi/7 - \cos \pi/21 + \cos 8\pi/21 = 1/2$,
$\cos \pi/7 - \cos 2\pi/7 + \cos 2\pi/21 - \cos 5\pi/21 = 1/2$,
$-\cos 2\pi/7 + \cos 3\pi/7 + \cos 4\pi/21 + \cos 10\pi/21 = 1/2$,
$-\cos \pi/15 + \cos 2\pi/15 + \cos 4\pi/15 - \cos 7\pi/15 = 1/2$.

In the current problem, we want to find the conditions for $\cos\left(\dfrac{2m_1 + 1}{N}\pi\right) + \cos\left(\dfrac{2m_3 + 1}{N}\pi\right) - 2\cos\left(\dfrac{2m_2 + 1}{N}\pi\right) = 0$, which for the moment can be written as $\cos \theta_1 + \cos \theta_3 - 2\cos \theta_2 = 0$, where $0 < \theta_i \leq \pi$. Before we can use CJ7, we need to consider the cases in which its assumptions are not satisfied. First, consider the case when $\theta_1 = \pi$ or $\theta_2 = \pi$. If $\theta_1 = \pi$, then $\cos \theta_3 - 2\cos \theta_2 = 1$, which is not possible by CJ7 if none of $\theta_2$ and $\theta_3$ is $\pi/2$. If indeed we allow $\theta_3 = \pi/2$ and $\theta_2 = 2\pi/3$, the equation is satisfied. However, in our original problem $\theta_i = \cos\left(\dfrac{2m_i+1}{N}\pi\right)$, therefore there are no integers of $N$ and $m_i$ that can make both $\theta_3 = \pi/2$ and $\theta_1 = \pi$, so this case is eliminated. On the other hand, if $\theta_2 = \pi$, then $\cos\theta_1 + \cos\theta_3 = -2$, which is impossible since we require $f(m_1) > f(m_2) > f(m_3)$.

Next, we consider the case where none of the three angles is $\pi$. Again, we need to first exclude the cases where one of the angles is $\pi/2$. Suppose $\theta_1 = \pi/2$, then $\cos\theta_3 - 2\cos\theta_2 = 0$, which is impossible by CJ7. On the other hand, if $\theta_2 = \pi/2$, then $\cos\theta_1 + \cos\theta_3 = 0$, which is impossible by CJ7 if $\theta_1$ and $\theta_3$ are distinct, as defined in CJ7. If $\theta_1$ and $\theta_3$ are equally spaced from $\pi/2$, $\theta_1 - \pi/2 = \pi/2 - \theta_3$, they will be considered not distinct by CJ7 and $\cos\theta_1 + \cos\theta_3 = 0$. Note that this case is just the one we originally proposed in the proposition: $\dfrac{2m_2+1}{N} = \dfrac{1}{2}$ and $m_1 - m_2 = m_2 - m_3$.

Finally, suppose none of the three angles are either $\pi$ or $\pi/2$. Then, if they are all distinct as defined by CJ7, we know $\cos\theta_1 + \cos\theta_3 - 2\cos\theta_2$ cannot be rational, and, therefore, cannot be zero. If two of the three angles are not distinct because they are equidistant from $\pi/2$, then either $-2\cos\theta_2 = 0$ or $\cos\theta_1 + 3\cos\theta_3 = 0$, which is impossible by CJ7.

To summarize, for all the cases where CJ7 does not apply, we have shown that the equality $\cos\left(\dfrac{2m_1+1}{N}\pi\right) + \cos\left(\dfrac{2m_3+1}{N}\pi\right) - 2\cos\left(\dfrac{2m_2+1}{N}\pi\right) = 0$ is satisfied if, and only if, $\theta_2 = \pi/2$ and $\theta_1 - \pi/2 = \pi/2 - \theta_3$. For the cases where CJ7 applies, we have shown that the equality is never possible. We conclude that the original proposition is true, giving the ferry state condition as $N = 4l + 2$.

### 3. Detailed mathematical derivations for the conditions of the ferry states

The antenna system consists of a ring of $N$ identical two-level optical emitters coupled through nearest-neighbor dipole-dipole interactions, whose Hamiltonian is given by:

$$H_a = \omega \sum_{i=1}^{N} \sigma_i^+ \sigma_i^- + S \sum_{i=1}^{N} (\sigma_i^+ \sigma_{i+1}^- + \sigma_{i+1}^+ \sigma_i^-) \tag{2.1}$$

where $\hbar = 1$, $\omega$ is the site energy, $S$ is the coupling strength, and $\sigma_{N+1}^\pm = \sigma_1^\pm$. The use of the Pauli raising and lowering operators explicitly ensures that a single site cannot support more than one excitation. Under Jordan-Wigner transformation, the ring Hamiltonian becomes:

$$H_{JW} = \omega \hat{n} + S \sum_{j=1}^{N-1} \left(c_j^\dagger c_{j+1} + c_{j+1}^\dagger c_j\right) - S\left(c_N^\dagger c_1 + c_1^\dagger c_N\right) e^{i\pi\hat{n}} \tag{2.2}$$

whose solution is given by:

$$|\psi_{m_1 m_2 \ldots m_n}\rangle = C^+_{m_1} C^+_{m_2} \ldots C^+_{m_n} |0\rangle \qquad (2.3)$$

where $m_1 \neq m_2 \neq \ldots \neq m_n$ and

$$C^+_m = \frac{1}{\sqrt{N}} \sum_{j=1}^{N} e^{i\frac{2m\pi}{N}j} c^\dagger_j \qquad \text{if } n \text{ is odd}$$

$$C^+_m = \frac{1}{\sqrt{N}} \sum_{j=1}^{N} e^{i\frac{(2m+1)\pi}{N}j} c^\dagger_j \qquad \text{if } n \text{ is even} \qquad (2.4)$$

are the creation operators for the component states. The energies of the n-excitation eigenstates are given by

$$\varepsilon_{m_1 m_2 \ldots m_n} = \sum_{i=1}^{n} \left(\omega + 2S \cos \frac{2m_i \pi}{N}\right) \qquad \text{if n is odd}$$

$$\varepsilon_{m_1 m_2 \ldots m_n} = \sum_{i=1}^{n} \left(\omega + 2S \cos \frac{(2m_i+1)\pi}{N}\right) \qquad \text{if n is even} \qquad (2.5)$$

where the total energy is obtained by summing over the energies of individual component states.

For single excitations, the eigenstates are simply given by:

$$|\psi_m\rangle = \frac{1}{\sqrt{N}} \sum_{j=1}^{N} e^{i\frac{2m\pi}{N}j} c^\dagger_j |0\rangle \qquad (2.6)$$

A general double-excitation state is:

$$\begin{aligned}
|\psi_{m_1 m_2}\rangle &= C^+_{m_1} C^+_{m_2} |0\rangle \\
&= \frac{1}{N} \sum_{j,h} e^{i\frac{(2m_1+1)\pi}{N}j} c^\dagger_j \cdot e^{i\frac{(2m_2+1)\pi}{N}h} c^\dagger_h |0\rangle \\
&= \frac{1}{N} \sum_{j,h} e^{i\frac{\pi}{N}[(2m_1+1)j+(2m_2+1)h]} c^\dagger_j c^\dagger_h |0\rangle \\
&= \frac{1}{N} \sum_{j<h} \left( e^{i\frac{\pi}{N}[(2m_1+1)j+(2m_2+1)h]} - e^{i\frac{\pi}{N}[(2m_1+1)h+(2m_2+1)j]} \right) \sigma^+_j \sigma^+_h |0\rangle
\end{aligned} \qquad (2.7)$$

where in the last line, we have carried out the Jordan-Wigner transformation. The coupling of $|\psi_{m_1 m_2}\rangle$ to a general single-excitation state is:

$$\Gamma = \left| \langle \psi_{m_1 m_2} | \sum_{j=1}^{N} \sigma_j^+ | \psi_k \rangle \right|^2$$

$$= \frac{1}{N} \left| \langle \psi_{m_1 m_2} | \sum_{j=1}^{N} \sigma_j^+ \sum_{h=1}^{N} e^{i\frac{2k\pi}{N}h} c_h^\dagger | 0 \rangle \right|^2$$

$$= \frac{1}{N} \left| \langle \psi_{m_1 m_2} | \sum_{j<h} \left( e^{i\frac{2k\pi}{N}h} + e^{i\frac{2k\pi}{N}j} \right) \sigma_j^+ \sigma_h^+ | 0 \rangle \right|^2 \quad (2.8)$$

$$= \frac{1}{N^3} \left| \sum_{j<h} \left( e^{i\frac{\pi}{N}[(2m_1+1)j+(2m_2+1)h]} - e^{i\frac{\pi}{N}[(2m_1+1)h+(2m_2+1)j]} \right) \left( e^{i\frac{2k\pi}{N}h} + e^{i\frac{2k\pi}{N}j} \right) \right|^2$$

First, consider the $|\psi_{m_1 m_2}\rangle$ states with even degeneracies that are formed by $|C_{m_1}\rangle$ and $|C_{m_2}\rangle$ of different energies. Their coupling to the bright state is calculated by setting $\frac{\pi}{N}[(2m_1+1)+(2m_2+1)] \neq 2\pi$ and $k=0$ in equation (9):

$$\Gamma_1 = \frac{4}{N^3} \left| \sum_{j<h} \left( e^{i\frac{\pi}{N}[(2m_1+1)j+(2m_2+1)h]} - e^{i\frac{\pi}{N}[(2m_1+1)h+(2m_2+1)j]} \right) \right|^2 = 0 \quad (2.9)$$

which is found to be zero by evaluating the first sum on the right:

$$\sum_{j<h} e^{i\frac{\pi}{N}[(2m_1+1)j+(2m_2+1)h]} = \sum_{s=1}^{N-1} \sum_{n=1}^{N-s} e^{i\frac{\pi}{N}[(m_1+m_2+1)2s+(2m_2+1)n]}$$

$$= 2 e^{i\frac{2(m_1+m_2+1)\pi}{N}} \cdot \left(1 - e^{i\frac{(2m_1+1)\pi}{N}}\right)^{-1} \cdot \left(1 - e^{i\frac{(2m_2+1)\pi}{N}}\right)^{-1} \quad (2.10)$$

where we find that the sum $\sum_{j<h} e^{i\frac{\pi}{N}[(2m_1+1)j+(2m_2+1)h]}$ is invariant with $m_1$ and $m_2$ switched, which leads to $\Gamma_1 = 0$ as in equation (10). Note the condition $\frac{\pi}{N}[(2m_1+1)+(2m_2+1)] \neq 2\pi$ has been used in evaluating the geometric series in equation (2.10). Indeed, if $\frac{\pi}{N}[(2m_1+1)+(2m_2+1)] = 2\pi$, the result in equation (2.10) will change to:

$$\sum_{j<h} e^{i\frac{\pi}{N}\left[(2m_1+1)j+(2m_2+1)h\right]} = \sum_{s=1}^{N-1}\sum_{n=1}^{N-s} e^{i\frac{\pi}{N}\left[(m_1+m_2+1)2s+(2m_2+1)n\right]}$$

$$= \sum_{s=1}^{N-1}\sum_{n=1}^{N-s} e^{i\frac{(2m_2+1)\pi}{N}n} \quad (2.11)$$

$$= \left(Ne^{i\frac{(2m_2+1)\pi}{N}} + 2 - N\right)\left(2 - 2\cos\frac{(2m_2+1)\pi}{N}\right)^{-1}$$

and by equation (10), the coupling strength will become:

$$\Gamma_2 = \frac{4}{N^3}\left|\frac{Ne^{i\frac{(2m_2+1)\pi}{N}} + 2 - N}{2 - 2\cos\frac{(2m_2+1)\pi}{N}} - \frac{Ne^{i\frac{(2m_1+1)\pi}{N}} + 2 - N}{2 - 2\cos\frac{(2m_1+1)\pi}{N}}\right|^2$$

$$= \frac{4}{N}\left(\frac{\sin\frac{(2m_2+1)\pi}{N}}{1-\cos\frac{(2m_2+1)\pi}{N}}\right)^2 \quad (2.12)$$

$$\neq 0$$

where the inequality to zero is obtained by noting $\frac{\pi}{N}\left[(2m_1+1)+(2m_2+1)\right] = 2\pi$ and $m_1 \neq m_2$, such that $\frac{(2m_2+1)\pi}{N} \neq \pi$. The physical interpretations of equations (10) and (11) are given in the main text. For the $|\psi_{m_1 m_2}\rangle$ states formed by $|C_{m_1}\rangle$ and $|C_{m_2}\rangle$ of the same energy as those considered in equations (2.11) and (11), the coupling is calculated by setting $\frac{\pi}{N}\left[(2m_1+1)+(2m_2+1)\right] = 2\pi$ and $k \neq 0$ in equation (9):

$$\Gamma_3 = \frac{1}{N^3}\left|\sum_{j<h}\left(e^{i\frac{\pi}{N}\left[(2m_1+1)j+(2m_2+1)h\right]} - e^{i\frac{\pi}{N}\left[(2m_1+1)h+(2m_2+1)j\right]}\right)\cdot\left(e^{i\frac{2k\pi}{N}h} + e^{i\frac{2k\pi}{N}j}\right)\right|^2$$

$$= \frac{1}{N^3}\left|\sum_{j<h}\left(e^{i\frac{\pi}{N}\left[-(2m_2+1)j+(2m_2+1)h\right]} - e^{i\frac{\pi}{N}\left[-(2m_2+1)h+(2m_2+1)j\right]}\right)\cdot\left(e^{i\frac{2k\pi}{N}h} + e^{i\frac{2k\pi}{N}j}\right)\right|^2 \quad (2.13)$$

$$= \frac{1}{N^3}\left|\sum_{j<h}\begin{pmatrix}e^{i\frac{\pi}{N}\left[-(2m_2+1)j+(2m_2+2k+1)h\right]} - e^{i\frac{\pi}{N}\left[-(2m_2+1)h+(2m_2+2k+1)j\right]} \\ +e^{i\frac{\pi}{N}\left[-(2m_2-2k+1)j+(2m_2+1)h\right]} - e^{i\frac{\pi}{N}\left[-(2m_2-2k+1)h+(2m_2+1)j\right]}\end{pmatrix}\right|^2$$

$$= 0$$

where the equality to zero is obtained by the same method in equation (2.10) – evaluating each term $\sum_{j<h} e^{i\frac{\pi}{N}(aj+bh)}$ on the right by converting it to:

$$\sum_{j<h} e^{i\frac{\pi}{N}(aj+bh)} = \sum_{s=1}^{N-1}\sum_{n=1}^{N-s} e^{i\frac{\pi}{N}[(a+b)s+bn]} = 2e^{i\frac{(a+b)\pi}{N}} \cdot \left(1-e^{i\frac{a\pi}{N}}\right)^{-1} \cdot \left(1-e^{i\frac{b\pi}{N}}\right)^{-1} \qquad (2.14)$$

which is invariant under a-b switch. Note that in equation (2.14), the condition $k \neq 0$ has been used when considering $a+b = -(2m_2+1)+(2m_2+2k+1) = -(2m_2-2k+1)+(2m_2+1) = 2k \neq 0$; hence, we can evaluate the geometric series as in equation (2.10), not as in equation (2.11). Equation (12) means that if a double-excitation state $|\psi_{m_1 m_2}\rangle$ is formed by $|C_{m_1}\rangle$ and $|C_{m_2}\rangle$ of the same energy, it will not couple to any dark state.

To prove that ferry states can always be constructed out of the double-excitation states with the accidental degeneracy, we consider the particular double-excitation states formed by $|C_{m_1}\rangle$ and $|C_{m_2}\rangle$ with $\frac{\pi}{N}[(2m_1+1)+(2m_2+1)] = \pi$:

$$\begin{aligned}
|\psi_{m_1 m_2}\rangle &= \frac{1}{N}\sum_{j<h}\left(e^{i\frac{\pi}{N}[(2m_1+1)j+(2m_2+1)h]} - e^{i\frac{\pi}{N}[(2m_1+1)h+(2m_2+1)j]}\right)\sigma_j^+\sigma_h^+|0\rangle \\
&= \frac{1}{N}\sum_{j<h}\left(e^{i\pi j}e^{i\frac{\pi}{N}(2m_2+1)(h-j)} - e^{i\pi h}e^{i\frac{\pi}{N}(2m_2+1)(j-h)}\right)\sigma_j^+\sigma_h^+|0\rangle
\end{aligned} \qquad (2.15)$$

and a particular dark state $|\psi_{m=N/2}\rangle = \frac{1}{\sqrt{N}}\sum_{j=1}^{N}(-1)^j c_j^\dagger |0\rangle$. The coupling between these two states is:

$$\begin{aligned}
\Gamma_4 &= \left|\langle\psi_{m_1 m_2}|\sum_{j=1}^{N}\sigma_j^+|\psi_{m=N/2}\rangle\right|^2 \\
&= \frac{1}{N}\left|\langle\psi_{m_1 m_2}|\sum_{j=1}^{N}\sigma_j^+\sum_{h=1}^{N}(-1)^h c_h^\dagger|0\rangle\right|^2 \\
&= \frac{1}{N}\left|\langle\psi_{m_1 m_2}|\sum_{j<h}\left((-1)^j + (-1)^h\right)\sigma_j^+\sigma_h^+|0\rangle\right|^2 \\
&= \frac{4}{N^3}\left|\sum_{j<h}\sin\frac{(2m_2+1)(h-j)\pi}{N}\left(1+(-1)^{j+h}\right)\right|^2
\end{aligned} \qquad (2.16)$$

Now, let $m_2 = 0$ and $m_1 = \frac{N-2}{2}$, such that $|\psi_{m_1 m_2}\rangle$ becomes $|\phi\rangle$ in the main text, we have

$$\Gamma_4 = \frac{4}{N^3} \left| \sum_{j<h} \sin\frac{(h-j)\pi}{N} \left(1+(-1)^{j+h}\right) \right|^2 \neq 0 \qquad (2.17)$$

where the inequality holds, since $0 < \frac{(h-j)\pi}{N} < \pi$ such that $\sin\frac{(h-j)\pi}{N} > 0$.

## 4. Phononic dissipation and disorder

As discussed in the main text, the ferry state effect can be observed under weak phononic dissipation, small $\omega$ disorder, and large $S$ disorder.

Under weak phononic dissipation, when $S < 0$, the ferry state effect is observable.

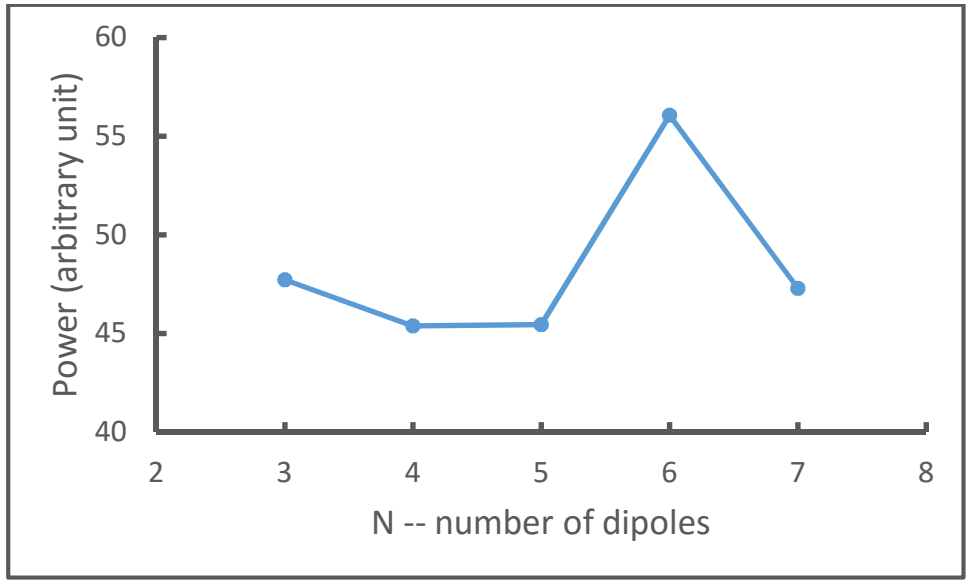

Figure 4. Ferry state effect under a weak phononic dissipation rate of $\gamma_p = 0.01\gamma_0$ with $S = -0.2eV$. All other parameters are as described in Table 1.

In Figure 4, the ferry state effect is clearly visible. Note that $S$ is negative here, causing the intraband transitions to favor the bright state, thereby making the ferry state effect more important in transferring excitation from the bright state to the dark states through interband transitions. If $\gamma_p = 0$ as discussed in the main text, the sign of $S$ does not matter since the only way to connect the bright state to the dark states is through the ferry states.

When there is disorder in $S$, we study the single-excitation eigenstates of the Jordan-Wigner Hamiltonian in equation (2.2) by having $n=2$, since we are interested in how the energy ladder of the component states changes with disorder. The general Hamiltonian with arbitrary disorder has the matrix form:

$$H = \begin{pmatrix} \omega & S_1 & 0 & . & . & S_N \\ S_1 & \omega & S_2 & & & \\ 0 & S_2 & . & & & \\ . & & & . & & \\ . & & & & . & S_{N-1} \\ S_N & & & & S_{N-1} & \omega \end{pmatrix} \qquad (3.1)$$

Assuming $N = 4l + 2$, the characteristic equation of this matrix is:

$$Det(H - \varepsilon \mathbf{I}) = \lambda Det(\mathbf{M}_{N-1}) + a Det(\mathbf{M}_{N-2}) + b = 0 \qquad (3.2)$$

where $\lambda = \omega - \varepsilon$ is the unknown of the equation, $\mathbf{M}_k$ is a rank $k$ tridiagonal matrix with $\lambda$'s only appearing on the diagonal, and $a$ and $b$ are coefficients only involving the $S_j$ variables. With mathematical induction, it is easy to prove that the determinant of the $\mathbf{M}_k$ type matrices only contains powers of $\lambda$ of the same parity of $k$:

$$\begin{aligned} Det(\mathbf{M}_k) &= \lambda^k + a_{k-2}\lambda^{k-2} + \cdots + a_0 & \text{if } k \text{ is even} \\ Det(\mathbf{M}_k) &= \lambda^k + a_{k-2}\lambda^{k-2} + \cdots + a_1\lambda & \text{if } k \text{ is odd} \end{aligned} \qquad (3.3)$$

Equation (3.3) implies the characteristic equation (3.2) can be written as:

$$\lambda Det(\mathbf{M}_{N-1}) + a Det(\mathbf{M}_{N-2}) + b = \lambda^N + a_{N-2}\lambda^{N-2} + \cdots + a_0 = 0 \qquad (3.4)$$

Where, since $N = 4l + 2$ is even, only the even powers of $\lambda$ appear. This means we can solve equation (3.4) by first finding solutions for $\lambda^2 = \lambda_1^2, \lambda_2^2, \lambda_3^2 \ldots \lambda_{N/2}^2$ and then take the square roots to get all the solutions $\lambda = \pm\lambda_1, \pm\lambda_2, \pm\lambda_3 \ldots \pm\lambda_{N/2}$. Since $\lambda = \omega - \varepsilon$, we see the eigenenergies of the component states are equally split around $\omega$. When there is no disorder, $S_1 = S_2 = \cdots = S_{N-1} = -S_N$, we find that some of the $\lambda^2$ solutions are equal and one specific $\lambda^2$ is equal to zero. In such a case, it is easy to identify the component states for the ferry state of the first category are states having finite equal splitting around $\omega$ and the component states for the ferry state of the second category are states having the same energy at $\omega$. Examining how the component states change under small perturbation of disorder, we see the component states for the first category still have equal splitting around $\omega$ (the splitting magnitudes are changed), and the component states for the second category are no longer degenerate, acquiring a small equal splitting around $\omega$. Consequently, the double-excitation ferry states are still degenerate with disorder as their energies are the sum of the component states. Certainly, with the perturbation of disorder, the characters of the component states are changed, so the ferry state effect will be diminished when the disorder increases. However, the preservation of the ferry state degeneracy is remarkable since it is proven with totally arbitrary choices for the coupling strength. Since the degeneracy is essential to the ferry state mechanism, we expect the ferry state effect is observable with relatively large disorder in $S$, as shown in Figure 5:

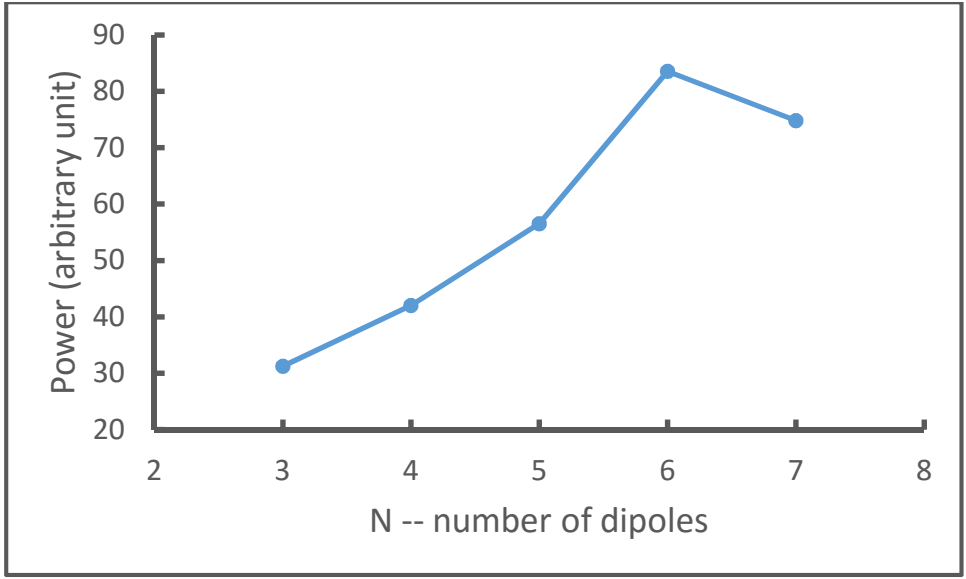

Figure 5. Ferry state effect under disorder in $S$. We assume a Gaussian distribution for $S$ with a mean of $\mu_S = 0.1 eV$ and a standard deviation of $0.05\mu_S$. All other parameters are as described in Table 1.

When there is disorder in the site energy $\omega$, the perturbation to the Hamiltonian in equation (2.2) is $V = \sum_{j=1}^{N} \delta_j c_j^+ c_j^-$. The unperturbed component states for the ferry states of the first category are a pair of $|C_{m_1}\rangle$ and $|C_{m_2}\rangle$ of different energies with $m_1 = 0$ and $m_2 = \frac{N-2}{2}$. Both $|C_{m_1}\rangle$ and $|C_{m_2}\rangle$ have their own degenerate counterpart and, to find the first order perturbation energy, we need to use the degenerate perturbation theory and diagonalize $P_{m_i} V P_{m_i}$, where $P_{m_i}$ is the projector into the degenerate space of $|C_{m_i}\rangle$. The result is a first order shift of $\alpha \pm \sqrt{|\beta|^2}$ for both the $|C_{m_1}\rangle$ and the $|C_{m_2}\rangle$ degenerate spaces, where $\alpha = \frac{1}{N}\sum_{j=1}^{N}\delta_j$ and $\beta = \frac{1}{N}\sum_{j=1}^{N}\delta_j e^{i\frac{2\pi j}{N}}$. Hence, by picking the eigenstate of energy $\varepsilon_{m_1} + \alpha + \sqrt{|\beta|^2}$ from the $|C_{m_1}\rangle$ degenerate space and the eigenstate of energy $\varepsilon_{m_2} + \alpha - \sqrt{|\beta|^2}$ from the $|C_{m_2}\rangle$ degenerate space, we can form a double-excitation state of the first category with the first order energy $\varepsilon_{m_1} + \varepsilon_{m_2} + 2\alpha$. The unperturbed component states for the ferry states of the second category are a pair of $|C_{m_3}\rangle$ and $|C_{m_4}\rangle$ of the same energy with $\frac{(2m_3+1)\pi}{N} = \frac{\pi}{2}$ and $\frac{(2m_4+1)\pi}{N} = \frac{3\pi}{2}$. Again, we use the perturbation theory for degenerate states and diagonalize $P_{m_3}VP_{m_3}$. The result is an energy shift of $\alpha \pm \gamma$ where $\gamma = \frac{1}{N}\sum_{j=1}^{N}\delta_j(-1)^j$, therefore, the double-excitation state formed by the eigenstates of $P_{m_3}VP_{m_3}$ has the first order energy of $2\varepsilon_{m_3} + 2\alpha$. Since $2\varepsilon_{m_3} = \varepsilon_{m_1} + \varepsilon_{m_2}$, we see the ferry state degeneracy is preserved up to first order. The second order energy, however, does not preserve the degeneracy and, therefore, the ferry state effect is less robust with disorder in the site energy than with disorder in the coupling strength. The numerical result is shown in Figure 6.

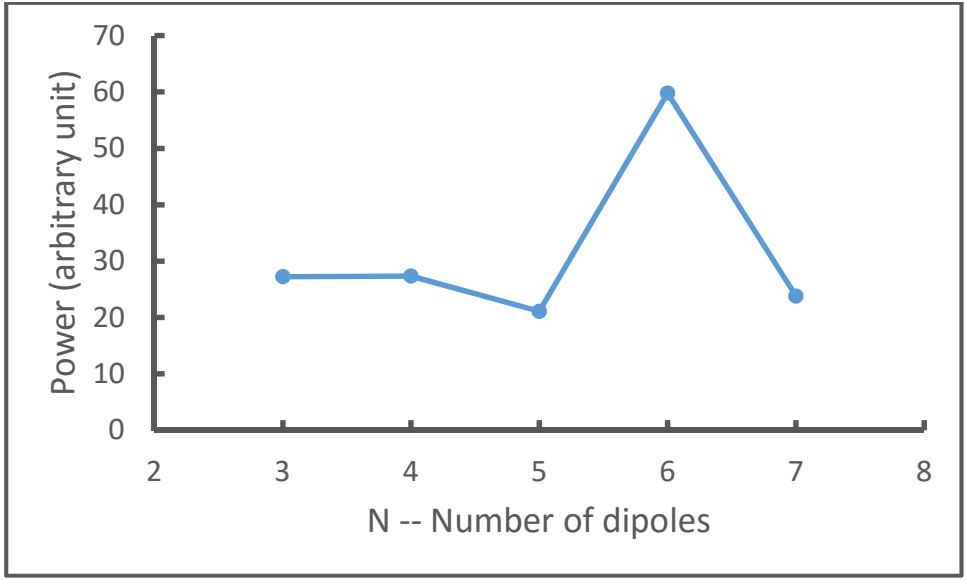

*Figure 6. Ferry state effect under disorder in $\omega$. We assume a Gaussian distribution for $\omega$ with a mean of $\mu_\omega = 1.76eV$ and a standard deviation of $0.001\mu_\omega$. All other parameters are as described in Table 1.*

To summarize this section, the ferry state effect can be observed under weak phononic dissipation, small $\omega$ disorder, and large $S$ disorder.

**Supplementary References:**